\documentclass{iau}
\usepackage{graphicx} 

\title[IAUS291.~~Superfluidity and entrainment in neutron star crusts] 
{How ``free'' are free neutrons in neutron-star crusts and what does it imply for pulsar glitches ?} 

\author[N. Chamel]  
{N. Chamel}

\affiliation{Institute of Astronomy and Astrophysics, Universit\'e Libre de Bruxelles, CP 226, Boulevard du Triomphe, B-1050 Brussels, 
Belgium\\[\affilskip]}

\pubyear{2012}
\volume{291}  
\jname{\mbox{Neutron Stars and Pulsars: Challenges and Opportunities after 80 years}}
\editors{J. van Leeuwen, ed.} 
\begin{document}

\maketitle

\begin{abstract}
The neutron superfluid permeating the inner crust of mature neutron stars is expected to play a key role in various astrophysical 
phenomena like pulsar glitches. Despite the absence of viscous drag, the neutron superfluid can still be coupled to the solid crust 
due to non-dissipative entrainment effects. Entrainment challenges the interpretation of pulsar glitches and suggests that a revision 
of the interpretation of other observed neutron-star phenomena might be necessary.
\keywords{stars: neutron, dense matter, equation of state, gravitation, hydrodynamics, stellar dynamics, (stars:) pulsars: general, stars: rotation}
\end{abstract}


\firstsection 
\section{Introduction}

Pulsars are among the most accurate clocks in the universe, the delays associated with their spin-down being at most of a few  
milliseconds per year. 
Nevertheless, some pulsars have been found to exhibit sudden increases in their rotational frequency $\Omega$. These 
``glitches'', whose amplitude varies from $\Delta\Omega\slash \Omega\sim 10^{-9}$ up to $\sim 10^{-5}$
(see e.g. Section 12.4 in \cite{lrr}). 
The long relaxation times following the first observed glitches and the glitches themselves hinted at the presence of superfluids in neutron-star 
interiors(\cite{bay69,pac72}). In fact, neutron-star superfluidity had been predicted by \cite{mig59} based on the 
microscopic theory of superconductivity developed by Bardeen, Cooper and Schrieffer two years earlier. Subsequently \cite{gk64} 
estimated the critical temperature for neutron superfluidity and suggested that the interior of a neutron star 
could be threaded by an array of quantized vortices. \cite{and75} advanced the seminal idea that pulsar glitches are 
triggered by the sudden unpinning of such vortices in the neutron-star crust. 
Their scenario found some support from laboratory experiments in superfluid helium (\cite{cam79,tsa80}). Further developments aimed at 
explaining the postglitch relaxation by the motion of vortices (\cite{alp85,jon93}). In the meantime, \cite{alp84} argued that the core of a 
neutron star (supposed to contain superfluid neutrons and superconducting protons) is unlikely to play any role in glitch 
events. Large pulsar glitches are still usually interpreted as sudden transfers of angular momentum between the neutron 
superfluid in the crust and the rest of the star. The confidence in this 
interpretation comes from i) the regularity observed in many glitching pulsars and ii) the fact that the estimated ratio of the moment 
of inertia $I_s$ of the superfluid component driving glitches to the total stellar moment of inertia $I$ is about $I_s/I\sim 1-2\%$ at 
most, as expected if only the crustal superfluid is involved (\cite{lnk99}). 

\section{Crustal entrainment}

Even though the neutron superfluid in the crust can flow without friction, it can still be entrained by nuclei, as 
first shown by \cite{carter05}. Indeed, unbound or ``dripped'' neutrons can be Bragg reflected by the crustal lattice 
in which case they cannot propagate and are therefore trapped in the crust. Unlike viscous drag, this entrainment effect 
is non-dissipative. 
Neutron diffraction is a well-known phenomenon, which has been routinely exploited to probe the structure of materials.  
Unlike the neutron beams used in terrestrial experiments, neutrons in neutron-star crusts are highly degenerate. Due to the 
Pauli exclusion principle, they must all have different (Bloch) wave vectors. As a result, they are simultaneously scattered 
in different directions. The strength of entrainment is therefore determined by the way all unbound neutrons are diffracted. 
This can be characterized by the density $n_n^{\rm c}$ of conduction neutrons, i.e. neutrons that are effectively ``free'' to move 
with a different velocity than that of nuclei. Equivalently, entrainment effects can be embedded in an effective neutron mass 
$m_n^\star=m_n n_n^{\rm f}/n_n^{\rm c}$ where $m_n$ is the bare neutron mass and $n_n^{\rm f}$ the density of unbound neutrons. 
Neutron conduction has been systematically studied in all regions of the inner crust using the band theory of solids (see \cite{cha12}). 
The neutron superfluid has thus been found to be very strongly entrained by the crust, especially in the region with
average baryon densities $\bar n\sim 0.02-0.03$~fm$^{-3}$. 

\section{Implications for pulsar glitches}

\begin{figure}[b]
\begin{center}
 \includegraphics[width=3.4in]{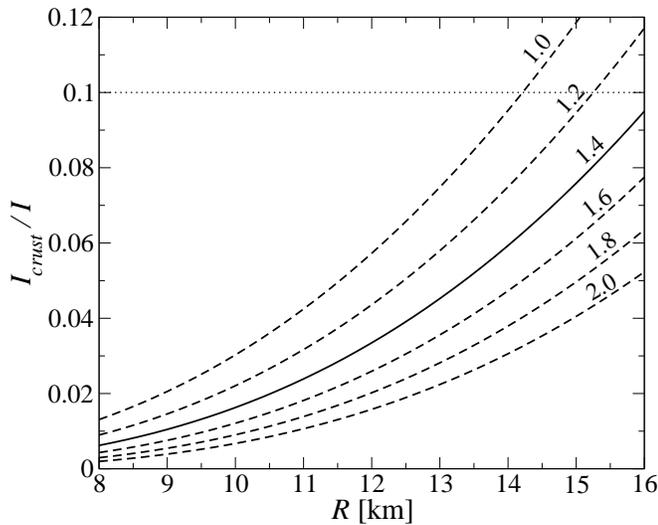} 
 \caption{Fractional moment of inertia of neutron-star crusts for different neutron-star masses (in solar masses) and radii. 
The horizontal dotted line indicates the lowest value consistent with Vela pulsar glitches.} 
   \label{fig1}
\end{center}
\end{figure}

According to a popular interpretation, pulsar glitches are due to sudden transfers of angular momentum between the neutron 
superfluid permeating the crust and the rest of the star. Due to entrainment, the angular momentum $J_{\rm s}$ of the superfluid depends 
not only on the angular 
velocity $\Omega_{\rm s}$ of the superfluid, but also on the (observed) angular velocity $\Omega$ of the star 
and can be expressed as (see \cite{cc06})
\begin{equation}
\label{1}
J_{\rm s}=I_{\rm ss} \Omega_{\rm s} + (I_{\rm s}-I_{\rm ss}) \Omega\, ,
\end{equation}
\cite[Chamel \& Carter (2006)]{cc06} showed that the product of the fractional moments of inertia $I_{\rm s}/I$ and $I_{\rm s}/I_{\rm ss}$ 
should obey the following constraint 
\begin{equation}
\label{2}
\frac{(I_{\rm s})^2}{I I_{\rm ss}}\geq \mathcal{G}\, , \hskip1cm \mathcal{G}\equiv \frac{1}{t}\sum_i\frac{\Delta\Omega_i}{|\dot\Omega|} \, ,
\end{equation}
where the sum is over all glitches observed during the time $t$ and $\dot\Omega$ is the observed average pulsar 
spin-down rate. A statistical study of glitching pulsars leads to $\mathcal{G}\simeq 1.7\%$ (see \cite{ly00}). 
If entrainment is 
neglected, $I_{\rm ss}=I_{\rm s}$ so that $J_{\rm s}=I_{\rm s} \Omega_{\rm s}$ and $(I_{\rm s})^2/(I I_{\rm ss})$ reduces to $I_{\rm s}/I$. 
Approximating this ratio by 
the fractional moment of inertia of the crust $I_{\rm crust}/I$, the constraint~(\ref{2}) was found to be easily satisfied for any 
realistic equation of state yielding plausible values for the neutron-star mass $M$ and radius $R$ (see \cite{lnk99}). On the other hand, 
taking entrainment into taken account (see ~\cite{cha12b}), we have found that $(I_{\rm s})^2/(I I_{\rm ss})\simeq 0.17 I_{\rm crust}/I$. 
Observations of large pulsar glitches and (\ref{2}) require that the fractional moment of inertia of the crust exceed $\sim 10\%$. 
The ratio $I_{\rm crust}/I$ can be estimated using the approximate expression 
of \cite{lat00}. 
The resulting fractional moment of inertia of the crust is shown in Fig.~\ref{fig1} for different neutron-star masses and radii using the 
neutron-star crust model of \cite{onsi08}.

This analysis implies that very active glitching pulsars should have an unusually small mass, significantly below the canonical value 
of $1.4M_\odot$ ($M_\odot$ being the mass of our Sun).
The existence of low mass neutron stars is not excluded, but such stars are not expected to be formed in a type II supernova 
explosion (see e.g. \cite{stro01}). Therefore, pulsars exhibiting large glitches should not be found in supernova remnants. This prediction 
is contradicted by the emblematic Vela pulsar, whose association with a supernova remnant is well established.

\section{Conclusions}

Due to entrainment effects, the neutron superfluid in neutron-star crusts does not carry enough angular momentum to explain 
large pulsar glitches. On the other hand, \cite{alp84} argued that the neutron superfluid in the core is also strongly coupled 
to the crust. The solution to this problem requires a closer examination of crustal entrainment and crust-core coupling. 
The presence of nuclear ``pastas'' at the crust bottom, quantum and thermal fluctuations of ions about their equilibrium 
positions, crystal defects, impurities and more generally any kind of disorder would presumably reduce the number of 
entrained neutrons. Further work is needed to confirm these espectations. However, \cite{sot12} have recently argued that 
quasiperiodic oscillation observed in giant flares from soft gamma-ray repeaters restrict the existence of pastas (if any) 
to a very narrow crustal region. Moreover, observations of the initial cooling in persistent soft X-ray transients are consistent 
with a low level of impurities in the crust (see \cite{sht07,bc09}) and this level is unlikely to be higher in non-accreting 
neutron stars like Vela. Incidentally these observations provided another proof for crustal superfluidity. 
On the other hand, the strong crust-core coupling assumed here could be much weaker, especially if protons form a type II
superconductor (see \cite{sed95b}). The observed rapid cooling of the neutron star in Cassiopeia A has 
recently provided strong evidence for core neutron superfluidity and proton superconductivity, but not on its type (see 
\cite{pag11,sht11}). \cite{lnk03} showed that type II superconductivity is incompatible with observations of long-period 
precession in pulsars. In the meantime, the work of \cite{ly10} has cast some doubt on the interpretation of long-period 
precession. In fact, proton superconductivity might be neither of type I nor of type II (see \cite{bab09}). In addition, neutron-star 
cores might contain other particle species with various superfluid and superconducting phases (see e.g. \cite{oer03}). This warrants
further studies. But if glitches are induced by the core of a neutron star, it will be challenging to explain the observed regularity of 
glitches and the fact that ${\cal G}\lesssim 2\%$. 

\acknowledgments This work was supported by FNRS (Belgium) and CompStar, a Research Networking Programme of the European 
Science Foundation. The author thanks B. Link and A. Alpar for valuable discussions.


\begin{thebibliography}{}
\bibitem[Alpar et al. (1984)]{alp84} Alpar, M.~A., Langer, S.~A. and Sauls, J.~A.\ 1984, \textit{Astrophys. J.}, 282, 533
\bibitem[Anderson \& Itoh (1975)]{and75} Anderson, P.~W. and Itoh, N. \ 1975, \textit{Nature}, 256, 25
\bibitem[Babaev 2009]{bab09}  Babaev, E.\ 2009, \textit{Phys. Rev. Lett.}, 103, 231101
\bibitem[Baym et al. 1969]{bay69} Baym, G., Pethick, C.~J. and Pines, D.\ 1969, \textit{Nature}, 224, 673
\bibitem[Brown \& Cumming 2009]{bc09} Brown, E.~F. and Cumming, A.\ 2009, \textit{Astrophys. J.}, 698, 1020
\bibitem[Campbell 1979]{cam79} Campbell, L.~J.\ 1979, \textit{Phys. Rev. Lett.}, 43, 1336 
\bibitem[Carter el al. (2005)]{carter05} Carter, B., Chamel, N. and Haensel, P. \ 2005, \textit{Nucl. Phys. A}, 748, 675
\bibitem[Chamel (2005)]{cha05} Chamel, N. \ 2005, \textit{Nucl.Phys. A}, 747, 109
\bibitem[Chamel \& Carter 2006]{cc06} Chamel, N. and Carter, B. \ 2006, \textit{Mon.Not.Roy.Astron.Soc.}, 368, 796 
\bibitem[Chamel \& Haensel 2008]{lrr} Chamel, N. and Haensel, P. \ 2008, ``Physics of Neutron Star Crusts'', 
\textit{Living Rev. Relativity} 11, 10. http://www.livingreviews.org/lrr-2008-10
\bibitem[Chamel 2012]{cha12} Chamel, N. \ 2012, \textit{Phys. Rev. C}, 85, 035801
\bibitem[Chamel 2013]{cha12b} Chamel, N. \ 2013, \textit{Phys. Rev. Lett.}, 110, 011101
\bibitem[Ginzburg \& Kirzhnits (1964)]{gk64} Ginzburg, V.~L. and Kirzhnits, D.~A. \ 1964, \textit{Zh. Eksp. Teor. Fiz.} 47, 2006
\bibitem[Jones 1993]{jon93} Jones, P.~B. \ 1993, \textit{Mon. Not. R. Astron. Soc.}, 263, 619 
\bibitem[Lattimer \& Prakash (2000)]{lat00} Lattimer, J.~M. and Prakash, M. \ 2000, \textit{Phys.Rep.}, 333, 121
\bibitem[Link et al. 1999]{lnk99} Link, B., Epstein, R.~I. and Lattimer, J.~M. \ 1999, \textit{Phys. Rev. Lett.}, 83, 3362
\bibitem[Link (2003)]{lnk03} Link, B. \ 2003, \textit{Phys. Rev. Lett.} 91, 101101
\bibitem[Lyne et al. 2000]{ly00} Lyne, A.~G., Shemar, S.~L., Smith, F.~G. \ 2000, \textit{Mon. Not. R. Astron. Soc.}, 315, 534
\bibitem[Lyne et al. (2010)]{ly10} Lyne, A.~G., Hobbs, G., Kramer, M., Stairs, I., Stappers, B. \ 2010, \textit{Science}, 329, 408
\bibitem[Migdal (1959)]{mig59} Migdal, A.~B. \ 1959, \textit{Nucl. Phys.}, 13, 655
\bibitem[Oertel \& Buballa 2006]{oer03} Oertel, M.  and Buballa, M. \ 2006, in: ``Color superconducting quark matter and the interior of neutron stars'' in 
Superdense QCD Matter and Compact Stars, ed. by D. Blaschke and D. Sedrakian. Proceedings of the NATO Advanced Research Workshop, 
27 September -- 4 October, 2003, in Yerevan, Armenia, Springer, Dordrecht, The Netherlands, p.187
\bibitem[Onsi et al. (2008)]{onsi08} Onsi, M., Dutta, A.~K., Chatri, H., Goriely, S., Chamel, N. and Pearson, J.~M. \ 2008, \textit{Phys. Rev. C}, 77, 065805
\bibitem[Packard 1972]{pac72} Packard, R. E. \ 1972, \textit{Phys. Rev. Lett.}, 28, 1080
\bibitem[Page et al. 2011]{pag11} Page, D., Prakash, M., Lattimer, J.~M., Steiner, A.~W. \ 2011, \textit{Phys. Rev. Lett.}, 106, 081101
\bibitem[Pines \& Alpar 1985]{alp85} Pines, D. and Alpar, M.~A. \ 1985, \textit{Nature}, 316, 27
\bibitem[Ruderman 1972]{rud72} Ruderman, M. \ 1972, \textit{Annu. Rev. Astron. Astrophys.}, 10, 427 
\bibitem[Sedrakian et al. 1995]{sed95b} Sedrakian, A.~D., Sedrakian, D.~M., Cordes, J.~M., Terzian, Y. \ 1995, \textit{Astrophys. J.}, 447, 324
\bibitem[Shternin, Yakovlev \& Haensel 2007]{sht07} Shternin, P.~S., Yakovlev, D.~G., Haensel, P., Potekhin, A.~Y. \ 2007, \textit{Mon.Not.Roy.Astron.Soc.}, 382, L43
\bibitem[Shternin 2011]{sht11} Shternin, P.~S., Yakovlev, D.~G., Heinke, C.~O., Ho, W.~C.~G., Patnaude, D.~J. \ 2011, \textit{Mon.Not.Roy.Astron.Soc.}, 412, L108
\bibitem[Sotani et al. (2012)]{sot12} Sotani, H., Nakazato, K., Iida, K., Oyamatsu, K. \ 2012, \textit{Phys. Rev. Lett.}, 108, 201101 
\bibitem[Tsakadze \& Tsakadze 1980]{tsa80} Tsakadze, J.~S.  and Tsakadze, S.~J. \ 1980, \textit{J. Low Temp. Phys.}, 39, 649 
\bibitem[Strobel \& Weigel 2001]{stro01} Strobel, K. and Weigel, M.~K. \ 2001, \textit{Astron. Astrophys.}, 367, 582
\end{thebibliography}
\end{document}